\documentclass[12pt]{article}

\catcode`\@=11
\@addtoreset{equation}{section}

\global\arraycolsep=2pt
\usepackage{amsbsy,amssymb,latexsym,amsfonts,amsmath,cite}

\newcommand\CL{{\mathcal L}}

\newcommand\CF{{\mathcal F}}
\newcommand\CM{{\mathcal M}}

\newcommand\CI{{\mathcal I}}
\newcommand\CJ{{\mathcal J}}

\newcommand\BM{{\boldsymbol{M}}}

\newcommand\nc{{noncommutative} }
\newcommand\Nc{{Noncommutative} }
\renewcommand\mod{~\text{mod}~}

\begin{document}

\hfill
KEK-TH-1076 \\
\hfill 
OIQP-06-03 \\ 
\hfill 
{\tt hep-th/0604039}

\vspace{10mm}

\begin{center}
{\Large \bf
Noncommutative D-branes
 from \\
 Covariant AdS Superstring
}
\end{center}
\vspace{10mm}

\centerline{\large Makoto Sakaguchi$^{a}$ and Kentaroh Yoshida$^{b}$}
\vspace{10mm}

\begin{center}
$^a$ {\it Okayama Institute for Quantum Physics\\
1-9-1 Kyoyama, Okayama 700-0015, Japan}\\
makoto$\_$sakaguchi@pref.okayama.jp
\vspace{4mm}

$^b${\it Theory Division, Institute of Particle and Nuclear Studies, \\
High Energy Accelerator Research Organization (KEK),\\
Tsukuba, Ibaraki 305-0801, Japan.} 
\\
kyoshida@post.kek.jp
\end{center}

\vfill

\begin{abstract}
We explore noncommutative D-branes in the AdS$_5\times$S$^5$ background
from the viewpoint of $\kappa$-invariance of a covariant open string
action in the Green-Schwarz formulation. 
Boundary conditions to ensure
the $\kappa$-invariance of the action lead to possible configurations of
noncommutative D-branes. With this covariant method, we 
derive 
configurations of 1/4 BPS noncommutative D-branes. The resulting D-branes other
than D-string are 1/4 BPS at any places, and the D-string is exceptional
and it is 1/2 BPS at the origin and 1/4 BPS outside the origin. All of
them are reduced to possible 1/4 BPS or 1/2 BPS AdS 
D-branes in the commutative limit or the
strong magnetic flux limit. We also apply the same analysis to an open
superstring in the pp-wave background and 
derive
configurations of 1/4 BPS noncommutative D-branes in the pp-wave. 
These D-branes consistently related to 
AdS D-branes
via the Penrose
 limit. 
\end{abstract} 

\thispagestyle{empty}
\setcounter{page}{0}

\newpage

\section{Introduction}

Noncommutative geometry in string theory has been well studied from the
novel work of Seiberg and Witten \cite{SW}.  When D-branes are
considered in the presence of a constant NS-NS two-form $B$, a low-energy
effective theory on the world-volume is realized as a field theory on a
noncommutative space. The constant two-form flux may be taken as a
magnetic flux on the world-volume. D-branes with constant fluxes (i.e.\
gauge field condensates on the branes) are often referred to as
noncommutative D-branes. Noncommutative D-branes in flat spacetime have
been well studied (for example, see \cite{CH,CZ,Rudychev}), but it seems not to
be the case for those on curved backgrounds. Here we are interested in
AdS-branes with gauge field condensates, which are interesting
objects to study in the context of the AdS/CFT correspondence \cite{AdS}.

An AdS-brane gives a defect in the field theory side, and another field
theory \cite{defect}, which is called defect conformal field theory
(dCFT), is realized on the defect \cite{dCFT}. When magnetic or electric
flux is turned on through the AdS-brane worldvolume, 
the flux should deform the dCFT and it
is interesting to study the resulting theory. For example, in the recent
study of the AdS/CFT duality at a far-from BPS region, an integrable open
spin chain appears in the analysis of the dCFT \cite{DM} \footnote{For
other open spin chains, see \cite{D7} for D3-D7-O7 and \cite{giant} for
a giant graviton.}. When the flux is turned on, it may be interpreted as a
boundary magnetic field of the open spin chain. For this purpose, as the
first step, we will study noncommutative D-branes in the
AdS$_5\times$S$^5$ and  pp-wave backgrounds \footnote{Noncommutative
D-branes in the pp-wave background are also discussed in \cite{CH-nc}}.

In a series of our previous works we showed the power of
$\kappa$-invariance of type IIB superstrings in a covariant
classification of possible D-branes in AdS$_5\times$S$^5$
\cite{Sakaguchi:2003py,Sakaguchi:2004md} (For a short review see
\cite{Sakaguchi:2004mj}) and in the pp-wave \cite{BPZ} \footnote{For a
classification of possible 1/2 BPS D-branes on a pp-wave with the
light-cone gauge, see \cite{DP,BP,SMT,Bain,BGG,SMT2}}. The
$\kappa$-invariance of an open string requires that surface terms under
the $\kappa$-variation should be deleted by imposing some additional
conditions \cite{LW}, which lead to the classification of the possible
D-branes. This procedure has some advantages in comparison to the brane
probe analysis with Dirac-Born-Infeld (DBI) action on the AdS and the
pp-wave \cite{SMT}. First, it is obvious that we do not need to analyze
each of DBI actions of D$p$-branes. All we have to consider is just a
string action. Secondly, our procedure is covariant and does not depend
on the light-cone gauge. Finally we can show that the classification is
valid at ``{\it full order}'' of the fermionic variables
\cite{Sakaguchi:2004md}. This procedure is also applicable to Dirichlet
branes for an open supermembrane on the AdS$_{4/7}\times$S$^{7/4}$
background \cite{Sakaguchi:2003hk,Sakaguchi:2004bu} and the pp-wave
\cite{Sakaguchi:2003ah,SY1}, and D-branes \cite{HPS} of type IIA pp-wave
string \cite{SY4}. These results are consistently related via the double
dimensional reduction \cite{DDR}.

\medskip 

In this paper we explore noncommutative D-branes in the
AdS$_5\times$S$^5$ background by analyzing a covariant type IIB string
action in the Green-Schwarz formulation \cite{MT-AdS}. We apply the
procedure developed in our previous papers to the case with a constant
two-form $\CF$ composed of $B$ and the field strength on the D-brane.
The boundary conditions to ensure that surface terms under the
$\kappa$-variation vanish are shown to lead to possible noncommutative
D-branes.\footnote{
In this paper, we consider surface terms up to the fourth order in $\theta$.
We expect that the result is valid even in the full order
as was proven in \cite{Sakaguchi:2004md}
for the commutative case.
}  As the result, we find 1/4 BPS noncommutative D-branes. The
D-branes other than D-string are 1/4 BPS anywhere, but the D-string is
exceptional and it is 1/2 BPS at the origin and 1/4 BPS outside the
origin. All of them are reduced to possible 1/4 BPS or 1/2 BPS AdS
D-branes in the commutative limit or the strong magnetic flux limit. We
also apply the same analysis to an open superstring in the pp-wave
background and derive configurations of 1/4 BPS noncommutative D-branes
in the pp-wave. These are consistently related to \nc AdS D-branes via
the Penrose limit \cite{Penrose}.

\medskip

This paper is organized as follows.  In section 2, we shall introduce a
covariant action of a type IIB open string in the AdS$_5\times$S$^5$
background in the Green-Schwarz formulation, which includes a constant
two-form $\CF$\,. Then we write down surface terms under the
$\kappa$-variation of the action, which play an important role in
determining possible configuration of noncommutative D-branes.  In
section 3, we discuss noncommutative D-branes in flat spacetime. The
well-known results obtained in the different methods can be reproduced
and this justifies the consistency of our method with the others.  In
section 4, we discuss 1/4 BPS noncommutative D-branes in
AdS$_5\times$S$^5$\,. In section 5, after introducing a covariant action
of a pp-wave superstring with a constant two-form, we find 1/4 BPS
noncommutative D-branes in the pp-wave. Section 6 is devoted to a
summary and discussions.

\section{Open AdS superstring with a constant two-form}

In this section we shall introduce the covariant action of the type IIB
string on the AdS$_5\times$S$^5$ background \cite{MT-AdS}, including a
constant two-form. The action has the $\kappa$-symmetry which ensures
the consistency of the theory. When we consider an open string rather
than a closed string, the $\kappa$-variation of the action gives surface
terms and so we need to impose some boundary conditions in order to
preserve the $\kappa$-symmetry \cite{LW}. These conditions restrict possible
configurations of D-branes in AdS$_5\times$S$^5$\,. 
We will derive the surface terms here, and the boundary
conditions will be considered in section 3 and 4.

\subsection{The action of AdS superstring}

The action of type IIB superstring on a supergravity background is given
by
\begin{eqnarray}
S&=&\int_\Sigma\! d^2\sigma\left[
\CL_{\rm NG}
+\CL_{\rm WZ}
\right]\,, 
\end{eqnarray}
where the Nambu-Goto (NG) part is \footnote{
We follow the notation used in \cite{Sakaguchi:2003py}.
} 
\begin{eqnarray}
\CL_{\rm NG}&=&
-\sqrt{-g(X,\theta)}~,\label{L:NG}\\
g_{ij}&=&E^M_iE^N_jG_{MN}=E^A_iE^B_j\eta_{AB}~,~~~
E^A_i=\partial_iZ^{\widehat M}E_{\widehat M}^A~,~~~
Z^{\widehat M}=(X^M,\theta^{\bar \alpha})~,
\end{eqnarray}
and the Wess-Zumino (WZ) part consists of the two parts \footnote{
For alternative superstring actions
in AdS$_5\times$S$^5$,
see \cite{alternativeND,alternative,RS} (see also
\cite{alternative2}). For classical integrability of AdS superstring
\cite{BPR} in \cite{RS}, see \cite{HY}. 
}: 
\begin{eqnarray}
\CL_{\rm WZ}&=&\CL_{\rm WZ}^0+\CL_{\rm WZ}^1~,\\
\CL_{\rm WZ}^1&=&-2i\int^1_0dt\widehat E^A\bar\theta\Gamma_A\sigma\widehat E~,\\
\CL_{\rm WZ}^0&=&
\frac{1}{2}\epsilon^{ij}e_i^Ae_j^B\CF_{AB}
=\frac{1}{2}\epsilon^{ij}\partial_iX^Me_M^A \partial_jX^Ne_N^B \CF_{AB}~,
\end{eqnarray}
where $\widehat E=E|_{\theta\to t\theta}$ and
 $\sigma=\sigma_3$ for an F-string while $\sigma=\sigma_1$ for a
D-string.  We will focus upon the F-string case hereafter. The two-from
field strength $\CF$ is defined by $\CF=B-da$ where $a$ is a gauge
field on the D-brane.

For the case of the AdS$_5\times$S$^5$ we are interested in, the
supervielbein
is given by  \cite{Sakaguchi:2003py}
\begin{eqnarray}
E^A = e^A + i\bar{\theta}\Gamma^A\left(
\frac{\sinh(\mathcal{M}/2)}{\mathcal{M}/2}
\right)^2D\theta\,, \qquad 
E^{\alpha} = 
\left(\frac{\sinh\mathcal{M}}{\mathcal{M}}D\theta\right)^{\alpha}
\,, 
\end{eqnarray}
where we have introduced the following quantities: 
\begin{eqnarray}
\mathcal{M}^2 &=& i\lambda\left(
\widehat{\Gamma}_Ai\sigma_2\theta\cdot\bar{\theta}\Gamma^A 
- \frac{1}{2}\Gamma_{AB}\theta\cdot \bar{\theta}\widehat{\Gamma}^{AB}i\sigma_2
\right)\,, \nonumber \\
D\theta &=& d\theta + \frac{\lambda}{2}e^A\widehat{\Gamma}_Ai\sigma_2\theta 
+ \frac{1}{4} \omega^{AB}\Gamma_{AB}\theta\,, 
\nonumber \\
\widehat{\Gamma}_A &\equiv& (-\Gamma_a\CI,\Gamma_{a'}\CJ)\,, \quad 
\widehat{\Gamma}_{AB} \;\equiv\;(-\Gamma_{ab}\CI,\Gamma_{a'b'}\CJ)\,,
\nonumber \\
\CI&=&\Gamma^{01234}~,~~~
\CJ=\Gamma^{56789}\,. 
\end{eqnarray}
Here $e$ and $\omega$ are the vielbein and the spin connection,
respectively, on the AdS$_5\times$S$^5$ background.

In the presence of a non-trivial constant $\CF$ along the D-brane world-volume
(i.e., $\CF_{\bar A\bar B}=const.$ and others $=0$), 
the boundary condition for Neumann directions is modified and becomes 
a mixed boundary condition. The resulting boundary condition is as
follows: 
\begin{eqnarray}
  \begin{array}{lll}
\partial_nX^Ne_N^{\bar {A}}
+\partial_tX^Ne_N^{\bar {B}}\CF_{\bar {B}}{}^{\bar {A}}=0~,       &&
\quad \bar {A}_i ~(i=0,\cdots,p)~ \in~ \mbox{Neumann}    \\
\partial_tX^Ne_N^{\underline{A}}=0~,      &&
\quad \underline{A}_i ~(i=p+1,\cdots,9)~ \in~ \mbox{Dirichlet}    \\
  \end{array}
\label{boundary condition}
\end{eqnarray}
where $\partial_t$ and $\partial_n$
are derivatives on $\partial\Sigma$
tangential
and normal 
to the D-brane world-volume, respectively.
In the limit for a magnetic flux, $\CF\to\infty$, Neumann directions
$(\bar A,\bar B)$ are replaced by Dirichlet directions.  On the other hand,
for an electric flux $F$, no dimensional reduction of the D-brane occurs    
because the range of the electric flux is restricted as $0\le\CF\le
1$\,.

\subsection{The $\kappa$-variation of the open string action}

According to our procedure explained in the introduction, let us examine
the surface terms under the $\kappa$-variation of the type IIB string
action on AdS$_5\times$S$^5$ up to fourth order in $\theta$\,.  For the
case with $\CF=0$, the vanishing conditions on the surface terms were
examined in \cite{Sakaguchi:2003py}.

To extract the surface terms of the fourth order in $\theta$ from
$S_{\rm WZ}^{\rm B} \equiv \int_\Sigma d^2\sigma\,\CL_{\rm WZ}^0$, we
need to know the expression of the $\kappa$-variation up to fourth order
in $\theta$.  The $\kappa$-variation $\delta_\kappa E^A=\delta_\kappa
Z^{\widehat M}E_{\widehat M}^A=0$ in this background is given as
\cite{Sakaguchi:2004md}
\begin{eqnarray}
\delta_\kappa X^Me_M^A=-(1+H+H^2+\cdots)^A{}_BE^B_{\bar {\alpha}}
\delta_\kappa\theta^{\bar {\alpha}}\,,
\end{eqnarray}
where
\begin{eqnarray}
H^A{}_B=-i\bar\theta\Gamma^A\left(
\frac{\sinh\CM/2}{\CM/2}
\right)^2[D\theta]_Ne_B^N~,~~~
E^A_{\bar {\alpha}}=i\bar\theta\Gamma^A
\left(
\frac{\sinh\CM/2}{\CM/2}
\right)^2~.
\end{eqnarray}
It is easy to derive
\begin{eqnarray}
\delta_\kappa X^Me_M^A&=&
-i\bar\theta\Gamma^A\delta_\kappa\theta
-\frac{i}{12}\bar\theta\Gamma^A\CM^2\delta_\kappa\theta
\nonumber\\&&
-\bar\theta\Gamma^A(\frac{\lambda}{2}\widehat\Gamma_Bi\sigma_2\theta
+\frac{1}{4}\omega_B{}^{CD}\Gamma_{CD}\theta)~
\bar\theta\Gamma^B\delta_\kappa\theta
+O(\theta^6)~.~~
\end{eqnarray}
By using this expression,
we obtain the $\kappa$-variation surface term of $S_{\rm WZ}^{\rm B}$
\begin{eqnarray}
\delta_\kappa S_{\rm WZ}^{\rm B} &=&
\int_{\partial\Sigma}d\xi\,\partial_tX^Me_M^A\CF_{AB}
\nonumber\\&&\times
\Bigg[
-i\bar\theta\Gamma^B\delta_\kappa\theta
-\frac{i}{12}\bar\theta\Gamma^B\CM^2\delta_\kappa\theta
\nonumber\\&&~~~
-\frac{\lambda}{2}\bar\theta\Gamma^B\widehat\Gamma_Ci\sigma_2\theta~
\bar\theta\Gamma^C\delta_\kappa\theta
-\frac{1}{4}\bar\theta\Gamma^B\Gamma_{CD}\theta~\omega_{E}^{CD}~\bar\theta\Gamma^E\delta_\kappa\theta
\Bigg]
+O(\theta^6)~.
\end{eqnarray}
It is known that the $\kappa$-variation of the total action $S$ gives
surface terms only. In particular the NG part produces no surface term,
and so the $\kappa$-variation surface terms emerge only from the WZ
part. 

The surface terms under the $\kappa$-variation of the WZ term 
are given by 
\begin{eqnarray}
\delta_\kappa S_{\rm WZ}
\big|
&=&
\int_{\partial\Sigma}d\xi\,
\Big[
\CL^0
+\CL^\lambda
+\CL^{\rm spin}
\Big]\,, 
\end{eqnarray}
with
\begin{eqnarray}
\CL^0&=&
-i\partial_tX^Me^A_M(\bar\theta\Gamma_A\sigma\delta_\kappa\theta
+\CF_{AB}\bar\theta\Gamma^B\delta_\kappa\theta
)
\nonumber\\&&
+\frac{1}{2}(
\bar\theta\Gamma^A\delta_\kappa\theta~
\bar\theta\Gamma_A\sigma\partial_t\theta
+
\bar\theta\Gamma_A\sigma\delta_\kappa\theta~
\bar\theta\Gamma^A\partial_t\theta
)
~,\\
\CL^\lambda&=&
\partial_tX^Me^A_M
\Big[
-\frac{\lambda}{2}(\bar\theta\Gamma_A\sigma\widehat\Gamma_C i\sigma_2\theta
+\CF_{AB}\bar\theta\Gamma^B\widehat\Gamma_C i\sigma_2\theta
)~
\bar\theta\Gamma^C\delta_\kappa\theta
\nonumber\\
&&
+\frac{\lambda}{4}(
\bar\theta\Gamma^B\delta_\kappa\theta~
\bar\theta\Gamma_B\widehat\Gamma_A\sigma i\sigma_2\theta
+
\bar\theta\Gamma^B\sigma\delta_\kappa\theta~
\bar\theta\Gamma_B\widehat\Gamma_Ai\sigma_2\theta
)\nonumber\\
&&
-\frac{i}{12}(\bar\theta\Gamma_A\sigma\CM^2\delta_\kappa\theta
+\CF_{AB}\bar\theta\Gamma^B\CM^2\delta_\kappa\theta
)
~\Big]~,\\
\CL^{\rm spin}&=&
-\frac{1}{4}\omega^{CD}_E\partial_tX^Me^A_M~
\bar\theta\Gamma^E\delta_\kappa\theta
(\bar\theta\Gamma_A\sigma\Gamma_{CD}\theta
+\CF_{AB}\bar\theta\Gamma^B\Gamma_{CD}\theta
)\nonumber\\
&&
+\frac{1}{8}\omega^{BC}_D\partial_tX^Me^D_M
(
\bar\theta\Gamma^A\delta_\kappa\theta~
\bar\theta\Gamma_A\Gamma_{BC}\sigma\theta
+
\bar\theta\Gamma^A\sigma\delta_\kappa\theta~
\bar\theta\Gamma_A\Gamma_{BC}\theta
)
~
\end{eqnarray}
where $\CL^{\rm spin}$ includes $\omega$-dependent terms, $\CL^\lambda$
includes $\lambda$-dependent (but $\omega$-independent) terms and
$\CL^0$ includes $\lambda$-independent terms. These surface terms
should vanish under some conditions to ensure the $\kappa$-invariance.
From the next section, we will examine the conditions under which the
surface terms vanish.

\section{Noncommutative D-branes in flat spacetime}

In this section we will examine the vanishing condition of
$\CL^0$\,. This condition leads to the classification of the possible
noncommutative D-branes in flat spacetime, since the AdS metric is
reduced to the flat spacetime metric when we consider the $\lambda=0$
case.

Noncommutative D-branes in the Green-Schwarz
formulation have been already studied
before \cite{CH,CZ,Rudychev},
but our analysis
here is covariant and so more general.   

First, let us examine the first line of $\CL^0$\,, which is 
the second order in $\theta$\,, 
\begin{eqnarray}
&&-i
\partial_tX^Me^{\bar A}_M(\bar\theta\Gamma_{\bar A}\sigma\delta_\kappa\theta
+\CF_{{\bar A}\bar B}\bar\theta\Gamma^{\bar B}\delta_\kappa\theta
)\nonumber\\
&&=
-i
\partial_tX^Me_M^{\bar A}\left[
(1+\CF)_{\bar {A}}{}^{\bar {B}}\bar\theta^1\Gamma_{\bar {B}}\delta_\kappa\theta^1
-(1-\CF)_{\bar {A}}{}^{\bar {B}}\bar\theta^2\Gamma_{\bar {B}}\delta_\kappa\theta^2
\right]~,
\label{second order in theta}
\end{eqnarray}
where we have used the boundary condition (\ref{boundary condition}).
We should carefully consider the gluing conditions for the fermionic
variable $\theta$ at boundaries in order to see that the surface terms
vanish. For this purpose we define the following matrix,
\begin{eqnarray}
M&\equiv&
g_0
\Gamma^{\bar {A}_0\cdots \bar {A}_p}
\prod_{n=1}^{[(p+1)/2]}(g_n+h_n\Gamma^{\bar a_{2n-1}\bar a_{2n}})~,~~~
\end{eqnarray}
and demand
\begin{eqnarray}
\theta^2=M\theta^1~.
\end{eqnarray}
Because $\theta^1$ and $\theta^2$ are Majorana-Weyl spinors,
$p$ must be odd, $p=1,3,5,\cdots$.

Let $C$ be the charge conjugation matrix satisfying 
\begin{eqnarray}
(\Gamma^A)^T=-C\Gamma^AC^{-1}\,, 
\end{eqnarray}
then one finds the following relation,
\begin{eqnarray}
\bar\theta^2&=&\bar\theta^1C^{-1}M^TC=\bar\theta^1M'~,\\
M'&\equiv&
(-1)^{p+1+[\frac{p+1}{2}]}g_0\prod_{n=1}(g_n-h_n\Gamma^{\bar a_{2n-1}\bar a_{2n}})
\Gamma^{\bar {A}_0\cdots \bar {A}_p}~.
\end{eqnarray}
If $\bar B\in \{\bar a_{2\ell-1},\bar a_{2\ell}\}$\,, then we find that 
\begin{eqnarray}
M'\Gamma_{\bar B}M&=&
-sg_0^2\prod(g_n^2+h_n^2s_n)
(g_\ell -h_\ell s_\ell
+2g_\ell h_\ell\epsilon_\ell)_{\bar B}{}^{\bar C}\Gamma_{\bar C}\,~
\end{eqnarray}
where $(\epsilon_\ell)_{\bar a_{2\ell-1}}{}^{\bar a_{2\ell}}= \eta^{\bar
a_{2\ell}\bar a_{2\ell}}$, $(\epsilon_\ell)_{\bar a_{2\ell}}{}^{\bar
a_{2\ell-1}}= -\eta^{\bar a_{2\ell-1}\bar a_{2\ell-1}}$ and others are
zero. Hence the vanishing condition of the first line of $\CL^0$ is
\begin{eqnarray}
(1+\CF)_{\bar A}{}^{\bar B}=
(1-\CF)_{\bar A}{}^{\bar C}
\left(
-sg_0^2
\prod_{n\neq \ell}(
g_n^2+h_n^2s_n
)
(g_\ell^2-h_\ell^2s_\ell
+2g_\ell h_\ell\epsilon_\ell)_{\bar C}{}^{\bar B}
\right)~.
\label{condition theta^2}
\end{eqnarray}

For the case without an electric $\CF$,
since
\begin{eqnarray}
\epsilon_\ell=\left(
  \begin{array}{cc}
    0   &1    \\
    -1   &0    \\
  \end{array}
\right)~,~~~
\left(
  \begin{array}{cc}
    0   &\CF_{\bar a_{2\ell-1}}{}^{\bar a_{2\ell}}    \\
    \CF_{\bar a_{2\ell}}{}^{\bar a_{2\ell-1}}    &0    \\
  \end{array}
\right)=\epsilon_\ell \CF_{\bar a_{2\ell-1}\bar a_{2\ell}}~
\end{eqnarray}
and $s_n=1$, (\ref{condition theta^2}) implies the following two equations: 
\begin{eqnarray}
1&=&
-sg_0^2\prod_{n\neq\ell}
(g_n^2+h_n^2)(g_\ell^2-h_\ell^2-2g_\ell h_\ell\CF\epsilon_\ell)
~,\label{eqn 1}\\
\CF&=&
-sg_0^2\prod_{n\neq\ell}
(g_n^2+h_n^2)(
2g_\ell h_\ell\epsilon_\ell
-(g_\ell^2-h_\ell^2)\CF)~,
\label{eqn 2}
\end{eqnarray}
for $1\le\ell\le[\frac{p+1}{2}]$\,. 
These are solved by
\begin{eqnarray}
g_0=\sqrt{-s}~,~~~
g_n=\cos\varphi_n~,~~~
h_n=\sin\varphi_n~,~~~
\CF_{\bar a_{2n-1}\bar a_{2n}}
=\tan\varphi_n
\end{eqnarray}
where $0\le \varphi_n\le\pi/2$\,.

For the case with an electric $\CF$\,,
say $\CF_{\bar a_1\bar a_2}$
and $\eta_{\bar a_1\bar a_1}=-\eta_{\bar a_2\bar a_2}=-1$\,,
since
\begin{eqnarray}
\epsilon_1=\left(
  \begin{array}{cc}
   0    &1    \\
   1    &0    \\
  \end{array}
\right)~,~~~
\left(
  \begin{array}{cc}
    0   &\CF_{\bar a_{1}}{}^{\bar a_{2}}    \\
    \CF_{\bar a_{2}}{}^{\bar a_{1}}    &0    \\
  \end{array}
\right)=\epsilon_1 \CF_{\bar a_{1}\bar a_{2}}~,
\end{eqnarray}
$s_1=-1$ and $s_n=1~(n\ge 2)$\,, 
(\ref{condition theta^2}) implies 
\begin{eqnarray}
1&=&
-sg_0^2\prod_{n\neq\ell}
(g_n^2+h_n^2s_n)(g_\ell^2-h_\ell^2s_\ell-2g_\ell h_\ell\CF\epsilon_\ell)
~,\label{eqn 1 ele}\\
\CF&=&
-sg_0^2\prod_{n\neq\ell}
(g_n^2+h_n^2s_n)(
2g_\ell h_\ell\epsilon_\ell
-(g_\ell^2-h_\ell^2s_\ell)\CF)~
\label{eqn 2 ele}
\end{eqnarray}
for $1\le\ell\le[\frac{p+1}{2}]$.
These are solved by
\begin{eqnarray}
&&
g_0=\sqrt{-s}~,
\\&&
g_1=\cosh\varphi_1~,~~~
h_1=\sinh\varphi_1~,~~~
\CF_{\bar a_{1}\bar a_{2}}
=\tanh\varphi_1
\\&&
g_n=\cos\varphi_n~,~~~
h_n=\sin\varphi_n~,~~~
\CF_{\bar a_{2n-1}\bar a_{2n}}
=\tan\varphi_n \quad (n\ge 2)\,, 
\end{eqnarray}
where $0\le \varphi_1\le\infty$ and $0\le \varphi_n\le\pi/2$~($n\ge
2$)\,. 

In both cases, $M$ is written as \footnote{ For the case with an
electric $\CF_{\bar a_1\bar a_2}$\,, we have used
$\cosh\varphi_1+\Gamma^{\bar a_{1}\bar a_{2}}\sinh\varphi_1
=\exp(\varphi_1\Gamma^{\bar a_{1}\bar a_{2}})$\,.  Because
$\varphi_1=\mathrm{arctanh}\, \CF_{\bar a_{1}\bar
a_{2}}=\frac{1}{2}\log\frac{1+\CF_{\bar a_{1}\bar a_{2}}}{1-\CF_{\bar
a_{1}\bar a_{2}}}$, we believe that the gluing matrix in the earlier
literature is valid only for an electric $\CF$\,.}
\begin{eqnarray}
M=\sqrt{-s}\exp\left(
\sum_{n=1}^{[(p+1)/2]}
\varphi_n\Gamma^{\bar a_{2n-1}\bar a_{2n}}
\right)
\Gamma^{\bar {A}_0\cdots \bar {A}_p}~.
\end{eqnarray}
It is clear from this expression that a \nc D$p$-brane is reduced to a
commutative D$p$-brane for $\forall \varphi_n\to 0$\,.  On the other
hand, it is reduced to a \nc D$(p-2)$-brane for a magnetic $\CF_{\bar
a_{2n-1}\bar a_{2n}}\to\infty$ ($\varphi_n\to\pi/2$)\,, while remains to
be a \nc D$p$-brane for an electric $\CF_{\bar a_1\bar a_2}\to 1$
($\varphi_1\to\infty$)\,.

It is convenient to rewrite the boundary condition $\theta^2=M\theta^1$
in the $2\times 2$ matrix form as follows:
\begin{eqnarray}
\theta=\BM\theta
=
\left(
  \begin{array}{cc}
    0   &M^{-1}    \\
    M   &0    \\
  \end{array}
\right)\left(
  \begin{array}{c}
  \theta^1     \\
  \theta^2     \\
  \end{array}
\right)~,~~~
\BM^2=1\,. 
\end{eqnarray}
Here the matrix $\BM$ is defined as 
\begin{eqnarray}
\BM
=\sqrt{-s}\exp(-\sum_{n=1}\varphi_n\Gamma^{\bar a_{2n-1}\bar a_{2n}}\sigma_3)
\Gamma\rho~
\label{M}
\end{eqnarray}
and then the matrix $\rho$ is represented by $2\times 2$ Pauli matrices,
according to the value of $p$\,, 
\begin{eqnarray}
\rho=\left\{
  \begin{array}{ll}
   \sigma_1    & \quad \text{when}~~p=1 \mod 4    \\
   -i\sigma_2    & \quad \text{when}~~p=3 \mod 4    \\
  \end{array}
\right.\,.
\end{eqnarray}
The range of the parameter $\varphi_n$ depends on the type of the flux,
namely magnetic or electric. For the magnetic flux case 
$\varphi_n$ takes the value in the range of $0\le \varphi_n\le\pi/2$\, 
and for the electric flux case it takes in $0\le \varphi_n\le\infty$\,.  
As we have just seen above, the gluing matrix $\BM$ satisfies
\begin{eqnarray} 
&&
\BM'\Gamma_{\bar A}\sigma_3+\Gamma_{\bar A}\sigma_3\BM
+\CF_{\bar A\bar B}(\BM'\Gamma^{\bar B}+\Gamma^{\bar B}\BM)=0~,
\label{key}
\end{eqnarray}
or equivalently
\begin{eqnarray}
\BM'(\Gamma_{\bar A}\sigma_3+\CF_{\bar A\bar B}\Gamma^{\bar B})
=-(\Gamma_{\bar A}\sigma_3+\CF_{\bar A\bar B}\Gamma^{\bar B})\BM
\label{key'}
\end{eqnarray}
where $\BM'$ is defined as 
\begin{eqnarray}
&&
\BM'=C^{-1}\BM^TC
=-\BM~.
\end{eqnarray}
It is obvious from (\ref{key'}) that
the first line of $\CL^0$ vanishes.

Then, by using (\ref{key}), the second line of $\CL^0$, the fourth order term
in $\theta$ vanishes as follows:
\begin{eqnarray}
&&
\frac{1}{2}(\bar\theta\Gamma^A\delta_\kappa\theta~
\bar\theta\Gamma_A\sigma
+
\bar\theta\Gamma^A\sigma\delta_\kappa\theta~
\bar\theta\Gamma_A)
\partial_t\theta
\nonumber\\&&
=
-\frac{1}{2}\CF_{\bar A\bar B}(\bar\theta\Gamma^{\bar A}\delta_\kappa\theta~
\bar\theta\Gamma^{\bar B}
+
\bar\theta\Gamma^{\bar B}\delta_\kappa\theta~
\bar\theta\Gamma^{\bar A})
\partial_t\theta
=0
~,
\end{eqnarray}
where we have used
\begin{eqnarray}
\bar\theta\Gamma^{\underline{A}}\delta_\kappa\theta
=-\bar\theta\BM\Gamma^{\underline{A}}\delta_\kappa\theta
=-\bar\theta\Gamma^{\underline{A}}\delta_\kappa\theta
=0~.
\end{eqnarray}

In summary we have discussed boundary conditions and
constructed the gluing matrix. As the result we have found the
well-known result that noncommutative D$p$-branes with $p=$ odd in flat
spacetime. By taking a strong flux limit $\CF\to \infty$ for a magnetic
case, D$p$-branes are reduced to D$(p-2)$-branes. The commutative limit
$\CF\to 0$ leads to the standard commutative D-branes.  The sequence of
them via the two limits is depicted in Table \ref{NC D flat}.
All of these D-branes are 1/2 BPS. In the case of flat spacetime higher
order terms than the fourth order in $\theta$ do not appear and so 
our result is rigorous with respect to $\theta$\,.

\begin{table}
\hspace*{2.7cm} {\footnotesize $\CF \to \infty$}  
\vspace*{-1.5cm}\\
\begin{center}
\begin{tabular}{cccccccccccccc}
& NC D9 & $\to$ & NC D7 & $\to$ & NC D5 & $\to$ & NC D3 & $\to$ & NC D1
 & $\to$ & C D$(-1)$   \\
 {\footnotesize $\CF \to 0$} 
&  $\downarrow$ & & $\downarrow$ & & $\downarrow$ & &
 $\downarrow$ & & $\downarrow$ &  & &   \\  
 & C D9 & &  C D7 & & C D5 & & C D3 & & C D1 & & 
\end{tabular} 
\caption{The sequence of noncommutative D-branes in flat spacetime.} 
\label{NC D flat}
\end{center}
\end{table}

\section{Noncommutative D-branes in  AdS$_5\times$S$^5$}

In this section, we examine $\lambda$-dependent terms $\CL^{\lambda}$
and $\omega$-dependent terms $\CL^{\rm spin}$. These terms include the
parameter $\lambda$\,, which characterizes the AdS geometry. Hence the
analysis here is intrinsic to the AdS case and the boundary conditions
lead to a classification of possible noncommutative AdS-branes. 

It is straightforward to see that
by using (\ref{key}),
$\CL^\lambda$ is rewritten as
\begin{eqnarray}
&\Bigg[&
\frac{\lambda}{2}
\bar\theta\Gamma^{\bar C}\delta_\kappa\theta~
\bar\theta(\Gamma_{\bar A}\sigma_3
+\CF_{\bar A\bar B}\Gamma^{\bar B})
\BM \widehat\Gamma_{\bar C}i\sigma_2
\theta~
\nonumber\\&&
+\frac{\lambda}{4}
\bar\theta\Gamma^{\bar B}\delta_\kappa\theta~
\bar\theta(\Gamma_{\bar B}\sigma_3
+\CF_{\bar B\bar C}\Gamma^{\bar C})
\BM \widehat\Gamma_{\bar A}i\sigma_2
\theta~\nonumber\\&&
+\frac{\lambda}{4}
\bar\theta\Gamma^{\underline{B}}\sigma_3\delta_\kappa\theta~
\bar\theta\Gamma_{\underline{B}}\widehat\Gamma_{\bar A}i\sigma_2\theta
\nonumber\\&&
+\frac{i}{12}
\bar\theta(\Gamma_{\bar A}\sigma_3
+\CF_{\bar {A}\bar {B}}
\Gamma^{\bar {B}})\BM\CM^2
\delta_\kappa\theta
~~
\Bigg]~\partial_tX^Me^{\bar A}_M~.
\label{forth order terms lambda-dependent}
\end{eqnarray}

The first and the second lines
in (\ref{forth order terms lambda-dependent}) vanish
when
\begin{eqnarray}
\BM \widehat\Gamma_{\bar C}i\sigma_2\theta
=\widehat\Gamma_{\bar C}i\sigma_2\theta\,, 
\label{cond 1}
\end{eqnarray}
which ensures that the third line  vanishes as
\begin{eqnarray}
\bar\theta\Gamma_{\underline{B}}\widehat\Gamma_{\bar A}i\sigma_2\theta
=-\bar\theta\Gamma_{\underline{B}}\BM\widehat\Gamma_{\bar A}i\sigma_2\theta
=-\bar\theta\Gamma_{\underline{B}}\widehat\Gamma_{\bar A}i\sigma_2\theta
=0~.
\end{eqnarray}

Among odd $p$, we find that this is satisfied for $p=1$\,. 
Noting that
\begin{eqnarray}
[\varphi_1\Gamma^{\bar a_1\bar a_2}\sigma_3,
\widehat\Gamma_{\bar C}i\sigma_2]=0
\end{eqnarray}
is satisfied by (2,0)- and (0,2)-branes, we find that (\ref{cond 1}) is
satisfied by the D1-branes. It is straightforward to see that the fourth
line vanishes for the D1-branes.  As we will see soon, $\CL^{\rm spin}$
vanishes for these branes sitting at the origin, and thus we find that
noncommutative D1-branes, (0,2) and (2,0), sitting at the origin are 1/2 BPS.

To proceed further, we shall introduce an additional gluing matrix
$\BM_0$ satisfying $\BM_0^2=1$\,, 
\begin{eqnarray}
\BM_0=\ell_0\Gamma^{\bar A_0\cdots \bar A_p} \,1~,~~~\ell=\left\{
  \begin{array}{ll}
    \sqrt{-s}   & \mbox{for}~p=1\mbox{~mod~}4    \\
    \sqrt{s}   & \mbox{for}~p=3\mbox{~mod~}4    \\
  \end{array}
\right.
\label{M_0}
\end{eqnarray}
and demand $\theta=\BM_0\theta$.
It satisfies
\begin{eqnarray}
\BM_0'=C^{-1}\BM_0^TC=(-1)^{p+1+[\frac{p+1}{2}]}\BM_0~.
\end{eqnarray}
The gluing matrices,
$\BM$ and $\BM_0$,
commute each other i.e., $[\BM,\BM_0]=0$\,, and
hence the branes obtained below are 1/4 BPS. 
Here we
should note that the most general 1/4 projections are not always written
in terms of mutually commuting gluing matrices only.  Hence 1/4 BPS
noncommutative D-branes we are considering is a part of possible 1/4 BPS 
D-branes.

By using the gluing condition $\theta=\BM_0\theta$\,, we can derive the
following equations: 
\begin{eqnarray}
&&\bar\theta\Gamma^{\bar C}\delta_\kappa\theta=
\bar\theta \BM_0'\Gamma^{\bar C}\BM_0\delta_\kappa\theta =0\,, 
\label{N M_0 AdS} 
\end{eqnarray}
when $p=3 \mod 4$\,, and 
\begin{eqnarray}
\bar\theta(\Gamma_{\bar A}\sigma_3
+\CF_{\bar A\bar B}\Gamma^{\bar B})
\BM \widehat\Gamma_{\bar C}i\sigma_2
\theta~
=0\,, 
\end{eqnarray}
when $p=1\,(3) \mod 4$ and $d,d'=\mbox{odd~(even)}$\,. 
Here $d\,(d')$ is the number of Dirichlet directions in AdS$_5$ (S$^5$), 
respectively. It follows from these equations that
the first and the second lines
in (\ref{forth order terms lambda-dependent})
vanish when one of the followings is satisfied
\begin{itemize}
  \item $p=3 \mod 4$,
  \item $p=1 \mod 4$ and $d=$ odd.
\end{itemize}
On the other hand, 
by noting that
\begin{eqnarray}
&&\bar\theta\Gamma^{\underline{C}}\sigma_3\delta_\kappa\theta=
0~~\mbox{when}~p=1 \mbox{~mod~}4~,
\label{M_0 D AdS}\\
&&\bar\theta\Gamma_{\bar B}\widehat\Gamma_{\bar A}i\sigma_2
=0
~~\mbox{when}~p=1\,(3) \mod 4~~\mbox{and}~~d,d'=\mbox{even\,(odd)}~,
\end{eqnarray}
we find the third line vanishes when 
one of the followings is satisfied
\begin{itemize}
  \item $p=1 \mod 4$,
  \item $p=3 \mod 4$ and $d,\,d'=$ odd.
\end{itemize} 
In summary we find that the first, the second and the third lines
vanish when $d=$odd and $p=1,3$ mod 4, namely 
we see that (even,even)-branes are possible.

Let us examine the fourth line.
Noting that (\ref{N M_0 AdS}) and
\begin{eqnarray}
&&
\bar\theta\Gamma^{\bar B}\BM\widehat\Gamma_{{A}}\theta
=0
~~~~
\text{when $A\in N(D)$, $p=1(3)\mod 4$ and $d,d'=$ odd},
\\&&
\bar\theta\Gamma^{\bar B}\BM\Gamma_{\bar A\underline{B}}\theta
=0
~~~~
\text{when $p=1\mod 4$},
\label{M_0 ND AdS}
\\&&
\bar\theta\Gamma^{\bar B}\BM\Gamma_{\bar A\bar{B}}\theta
=\bar\theta\Gamma^{\bar B}\BM\Gamma_{\underline{A}\underline{B}}
\delta_\kappa\theta
=
0
~~~~
\text{when $p=3\mod 4$},
\\&&
\bar\theta\widehat\Gamma^{\bar A\underline{B}}i\sigma_2\delta_\kappa\theta
=0
~~~~
\text{when $p=1(3)\mod 4$ and $d,d'=$ even(odd)},
\\&&
\bar\theta\Gamma^{\bar A\bar{B}}i\sigma_2\delta_\kappa\theta
=\bar\theta\Gamma^{\underline{A}\underline{B}}i\sigma_2\delta_\kappa\theta
=0
~~~~
\text{when $p=1(3)\mod 4$ and $d,d'=$ odd(even)}~,~~~~
\end{eqnarray}
we find the fourth line  vanishes 
for (even,even)-branes. 

Next, we examine $\omega$-dependent terms, $\CL^{\rm spin}$,
which 
are rewritten as
\begin{eqnarray}
&&
\frac{1}{4}\omega^{CD}_{\bar E}\partial_tX^Me^{\bar A}_M~
\bar\theta\Gamma^{\bar E}\delta_\kappa\theta~
\bar\theta(\Gamma_{\bar A}\sigma_3
+\CF_{\bar A\bar B}~\Gamma^{\bar B})
\BM\Gamma_{CD}\theta
\nonumber\\
&&
-\frac{1}{8}\omega^{BC}_{\bar D}\partial_tX^Me^{\bar D}_M~
\bar\theta\Gamma^{\bar A}\delta_\kappa\theta~
\bar\theta(\Gamma_{\bar A}\sigma_3
+\CF_{\bar A\bar E}~\Gamma^{\bar E})
\BM\Gamma_{BC}
\theta
\nonumber\\
&&
+\frac{1}{8}\omega^{BC}_{\bar D}\partial_tX^Me^{\bar D}_M~
\bar\theta\Gamma^{\underline{A}}\sigma_3\delta_\kappa\theta~
\bar\theta\Gamma_{\underline{A}}\Gamma_{BC}\theta
~.
\end{eqnarray}
Let us consider 1/2 BPS D1-branes, (2,0)- and (0,2)-branes, first.
The vanishing condition for the first and the second lines 
is
\begin{eqnarray}
\omega^{CD}_{\bar E}\BM\Gamma_{CD}\theta=
\omega^{CD}_{\bar E}\Gamma_{CD}\theta
\label{cond2}
\end{eqnarray}
which ensures that the third line vanishes.
Since $\omega^{\underline{C}\underline{D}}_{\bar E}=0$
(see Appendix B in \cite{Sakaguchi:2003py})
 and
$\BM\Gamma_{\bar C\bar D}=\Gamma_{\bar C\bar D}\BM$
for these branes,
the first and the second lines vanish if 
$\omega^{\bar{C}\underline{D}}_{\bar E}=0$,
i.e., if branes are sitting at the origin.
As $\bar\theta\Gamma_{\underline{A}}\Gamma_{\bar B\bar C}\theta=0$
for these branes,
the last line vanishes if 
$\omega^{\bar{B}\underline{C}}_{\bar D}=0$,
i.e., if the branes are sitting at the origin.
Thus (2,0)- and (0,2)-branes sitting at the origin are 1/2 BPS.

Next we consider 1/4 BPS D-branes.  First we examine the first and the
second lines.  They vanish when $p=3 \mod 4$ due to (\ref{N M_0 AdS})\,.
For $p=1\mod 4$\,, the terms proportional to $\omega^{\bar
C\underline{D}}$ vanish due to (\ref{M_0 ND AdS})\,.  Noting that
$\omega^{\underline{C}\underline{D}}_{\bar E}=0$\,, we are left with the
terms including $\omega^{\bar{C}\bar{D}}_{\bar E}$\,.  It is natural to
expect that $\omega^{\bar C\bar D}=0$ for $\CF_{\bar B\bar C}\neq 0\neq
\CF_{\bar D\bar E}$\,. This implies that (\ref{cond2}) is satisfied for
$p=1 \mod 4$.  Thus the first and the second lines vanish under the
configuration even outside the origin.  Next we examine the last line.
It vanishes for $p=1\mod 4$ due to (\ref{M_0 D AdS}).  Since
\begin{eqnarray}
\bar\theta\Gamma_{\underline{A}}\Gamma_{\bar B\underline{C}}\theta=0
~~~\text{when $p=3\mod 4$}
\end{eqnarray}
and $\omega^{\underline{B}\underline{C}}_{\bar D}=0$\,,
we are left with terms proportional to $\omega^{\bar B\bar C}$\,. 
By the same reasoning above, we derive
\begin{eqnarray}
\omega^{\bar B\bar C}_{\bar D}\bar\theta\Gamma_{\underline{A}}\Gamma_{\bar B\bar C}\theta
=\omega^{\bar B\bar C}_{\bar D}\bar\theta\BM'\Gamma_{\underline{A}}\Gamma_{\bar B\bar C}\theta
=-\omega^{\bar B\bar C}_{\bar D}\bar\theta\Gamma_{\underline{A}}\Gamma_{\bar B\bar C}\BM\theta
=0\,, 
\end{eqnarray}
so that the last line vanishes even for $p=3\mod 4$\,. 
We note that $\CL^{\rm spin}$ vanishes under the configuration
even outside the origin.

Summarizing, we find  1/4 BPS \nc  D-branes depicted in Table 
\ref{1/4 BPS D AdS}.
$(2,0)$- and $(0,2)$-branes are 1/2 BPS when they are sitting at the origin,
while 1/4 BPS when they move away from the origin.

\vspace*{0.5cm}
\begin{table}[htbp]
 \begin{center}
  \begin{tabular}{|c|c|c|c|c|}
    \hline
D1  & D3   & D5   & D7   &D9   \\
    \hline
(2,0), ~(0,2)   &(4,0), ~(0,4),~(2,2)    & (4,2),~ (2,4)   & (4,4)   & absent   \\
    \hline
  \end{tabular}
 \caption{1/4 BPS \nc  D-branes in AdS$_5\times$S$^5$} 
 \label{1/4 BPS D AdS}
 \end{center}
\end{table}

When $\forall\CF\to 0$, 1/4 BPS \nc D$p$-branes with $p=3$ mod 4
are reduced to 1/4 BPS commutative D$p$-branes,
while 1/4 BPS \nc D$p$-branes with $p=1$ mod 4
become  1/2 BPS commutative D$p$-branes because 
surface terms vanish without $\BM_0$
in this limit \cite{Sakaguchi:2003py}.
We summarize commutative D-branes
sitting at the origin in Table 3.

\vspace*{0.5cm}
\begin{table}[htbp]
 \begin{center}
  \begin{tabular}{|c|c|c|c|c|c|c|}
    \hline
SUSY&D($-1$)&D1  & D3   & D5   & D7   &D9   \\
    \hline
1/2&(0,0)&(2,0), ~(0,2)   &(3,1),~(1,3)   & (4,2),~ (2,4)   & (3,5),~(5,3)   & absent   \\
1/4&& &(4,0), ~(0,4),~(2,2)    &   & (4,4)   &    \\
    \hline
  \end{tabular}
 \caption{Commutative  D-branes in AdS$_5\times$S$^5$ sitting at the origin}
 \label{Com D AdS}
 \end{center}
\end{table}

\subsection*{Penrose limit}

Now let us discuss the Penrose limit \cite{Penrose} of the above branes.
The Penrose limit is taken as follows \cite{IW10}. First let us introduce the
light-cone coordinates as $X^\pm=\frac{1}{\sqrt{2}}(X^9\pm X^0)$ and
$\theta_\pm=P_{\pm}\theta_\pm$ with $P_\pm=\frac{1}{2}\Gamma_+\Gamma_-$,
and scale as
\begin{eqnarray}
X^+\to \Omega^2 X^+~,~~~
\theta_+\to \Omega \theta_+\,. 
\end{eqnarray}
Then the limit $\Omega\to0$ is taken. We distinguish the cases
depending on the boundary conditions of the light-cone coordinate
$(X^+,X^-)$ as (N,N) for $X^\pm \in$ Neumann directions and (D,D) for
$X^\pm \in$ Dirichlet directions. 
We consider the Penrose limits of (N,N)- and (D,D)-cases. 
Then we see that the noncommutative AdS-branes obtained above 
are reduced to 
noncommutative D-branes in the pp-wave as shown in Table \ref{Penrose:tab}. 

\vspace*{0.5cm}
\begin{table}[htbp]
\begin{center}
\begin{tabular}{ll}
$(4,0)_{\frac{1}{4}}\xleftarrow{\rm DD}
(4,0)_{\frac{1}{4}}\xrightarrow{\rm NN}\times$   
& \qquad 
$(0,4)_{\frac{1}{4}}\xleftarrow{\rm DD}
(0,4)_{\frac{1}{4}} 
\xrightarrow{\rm NN}\times$ 
\\ 
$(2,2)_{\frac{1}{4}}\xleftarrow{\rm DD}
(2,2)_{\frac{1}{4}}\xrightarrow{\rm NN}(+-;1,1)_{\frac{1}{4}}$  
& \qquad  
$(4,4)_{\frac{1}{4}}\xleftarrow{\rm DD}
(4,4)_{\frac{1}{4}}\xrightarrow{\rm NN}(+-;3,3)_{\frac{1}{4}}$  
\\
$(2,0)_{\frac{1}{4}}\xleftarrow{\rm DD}
(2,0)_{\frac{1}{4}}\xrightarrow{\rm NN}\times$ 
& \qquad 
$(0,2)_{\frac{1}{4}}\xleftarrow{\rm DD}
(0,2)_{\frac{1}{4}}\xrightarrow{\rm NN}\times$  
\\
$(4,2)_{\frac{1}{4}}\xleftarrow{\rm DD}
(4,2)_{\frac{1}{4}}\xrightarrow{\rm NN}(+-;3,1)_{\frac{1}{4}}$  
& \qquad 
$(2,4)_{\frac{1}{4}}\xleftarrow{\rm DD} 
(2,4)_{\frac{1}{4}}\xrightarrow{\rm NN}(+-;1,3)_{\frac{1}{4}}$  
\end{tabular}
\end{center}
\caption{Penrose limit of noncommutative AdS-branes.} 
\label{Penrose:tab}
\end{table}

In the subsequent sections, we will discuss noncommutative D-branes in
a pp-wave by studying the $\kappa$-invariance of a covariant string
action on the pp-wave. We will see that the possible noncommutative
D-branes are ($+-$;odd,odd)- and (even,even)-type configurations and all
of them are consistently obtained in the Penrose limit considered above.

\section{Noncommutative D-branes in the pp-wave}

In this section we shall consider an open superstring in the maximally
supersymmetric pp-wave background. D-branes in the pp-wave have been
well studied and noncommutative D-branes in the pp-wave are also
discussed in \cite{CH-nc}. However, most of the earlier works are done
in the light-cone gauge. As far as we know, there are no covariant
studies of noncommutative D-brane in the pp-wave. Hence, by repeating
the analysis in the case of the AdS superstring, let us classify
possible noncommutative D-branes in the pp-wave. This classification is
also our new result.

\subsection{Open pp-wave superstring with a constant two-form}

For the pp-wave background supervielbeins are given by 
\cite{Sakaguchi:2003py}
\begin{eqnarray*}
E^A=e^A+\frac{i}{2}\bar\theta\Gamma^A\left(\frac{\sinh(\CM/2)}{\CM/2}\right)^2
D\theta\,,
\quad 
E^{{\alpha}}=
\left(\frac{\sinh\CM}{\CM}D\theta\right)\,. 
\end{eqnarray*}
Here we have introduced several quantities: 
\begin{eqnarray*}
&& \CM^2 =
i\frac{\mu}{2}(f+g)i\sigma_2\theta\cdot \bar\theta\Gamma^-
+i\frac{\mu}{2}\widehat\Gamma_mi\sigma_2\theta\cdot \bar\theta\Gamma^m
+i\frac{\mu}{2}\Gamma_{r}\Gamma_+\theta\cdot \bar\theta\Gamma^{r}fi\sigma_2
\\&& \qquad \quad 
-i\frac{\mu}{2}\Gamma_{r'}\Gamma_+\theta\cdot \bar\theta\Gamma^{r'}gi\sigma_2
-i\frac{\mu}{2}{\Gamma}_{mn}\theta\cdot \bar\theta\widehat{\Gamma}^{mn}
i\sigma_2\,,\\
&& D\theta = d\theta
+e^-~\frac{\mu}{2}(f+g)i\sigma_2\theta
+e^m~\frac{\mu}{2}\widehat{\Gamma}_mi\sigma_2\theta
+e^m_*~\frac{\mu^2}{2}\Gamma_m\Gamma_+\theta\,,
\\
&& \widehat\Gamma_m=(-\Gamma_r\Gamma_+f,~\Gamma_{r'}\Gamma_+g)\,,~~~
\widehat\Gamma_{mn}=(-\Gamma_{rs}\Gamma_+f,~\Gamma_{r's'}\Gamma_+g)\,,
\\
&& e^+ = dX^+-\frac{\mu^2}{2}(X^m)^2dX^-\,,~~~
e^-=dX^-\,,~~~
e^m=dX^m\,,~~~
e^m_*=X^mdX^-\,.
\end{eqnarray*}
In this parametrization, the pp-wave metric becomes the standard form 
\begin{eqnarray}
ds^2=2e^+e^-+(e^m)^2
~=~2dX^+dX^--\mu^2(X^m)^2(dX^-)^2+(dX^m)^2\,.
\nonumber
\label{metric:PP}
\end{eqnarray} 
The WZ action in the case of the pp-wave is given by 
\begin{eqnarray}
S_{\rm WZ}=\int_\Sigma d^2\sigma\left[
\CL^0_{\rm WZ}+\CL_{\rm WZ}^1
\right]
\end{eqnarray}
with
\begin{eqnarray}
\CL^0_{\rm WZ}&=&\epsilon^{ij}\partial_iX^Me_M^A\partial_jX^Ne_N^B\CF_{AB}~,\\
\CL^1_{\rm WZ}&=&\epsilon^{ij}\Big[
-i\partial_iX^Me_M^A\bar\theta\Gamma_A\sigma D_j\theta
-\frac{i}{12}\partial_iX^Me_M^A\bar\theta\Gamma_A\sigma\CM^2D_j\theta
\nonumber\\&&~~~
+\frac{1}{4}\bar\theta\Gamma^AD_i\theta~
\bar\theta\Gamma_A\sigma D_j\theta
\Big]+O(\theta^6)~.
\end{eqnarray}

The universal feature of  $\kappa$-variation is 
\begin{eqnarray}
\delta_\kappa E^A=\delta_\kappa 
X^ME_M^A+\delta_\kappa\theta^\alpha E_\alpha^A=0 
\end{eqnarray}
and for the pp-wave case it can be rewritten, at the
fourth order of $\theta$\,, as 
\begin{eqnarray}
\delta_\kappa X^Me_M^A&=&
-\frac{i}{2}\bar\theta\Gamma^A\delta_\kappa\theta
-\frac{i}{24}\bar\theta\Gamma^A\CM^2\delta_\kappa\theta
-\frac{1}{4}\bar\theta\Gamma^B\delta_\kappa\theta~
\bar\theta\Gamma^A(\varrho_B+\varrho^{\rm spin}_B)\theta
+O(\theta^6)\,, 
\end{eqnarray}
where $\varrho_A$ and $\varrho_A^{\rm spin}$ are defined, respectively, by
\begin{eqnarray}
\varrho_A
&=&e^M_A\left(
e^-_M\frac{\mu}{2}(f+g)i\sigma_2
+e^m_M\frac{\mu}{2}\widehat\Gamma_mi\sigma_2
\right)\,, \quad 
\varrho^{\rm spin}_A
=
e^M_A
\left(
e^m_{*M}\frac{\mu^2}{2}\Gamma_m\Gamma_+
\right)\,. 
\end{eqnarray}
It is straightforward to see that
\begin{eqnarray}
\varrho_-=\frac{\mu}{2}(f+g)i\sigma_2\,, \quad 
\varrho_m=\frac{\mu}{2}\widehat\Gamma_{m}i\sigma_2\,, \quad 
\varrho_-^{\mathrm{spin}}=\frac{\mu^2}{2}X^m\Gamma_m\Gamma_+\,, 
\end{eqnarray}
and others vanish.

Under the $\kappa$-variation,
$S_{\rm NG}$ does not produce a surface term.
The surface terms under the $\kappa$-variation of $S_{\rm WZ}$ are obtained as
\begin{eqnarray}
\delta_\kappa S_{\rm WZ}\big|
&=&\int_{\partial\Sigma}d\xi\left[
\CL^0
+\CL^\mu
+\CL^{\rm spin}
\right]~
\end{eqnarray}
with
\begin{eqnarray}
\CL^0&=&
-i\left(\bar\theta\Gamma_A\sigma\delta_\kappa\theta
+
\CF_{AB}\bar\theta\Gamma^B\delta_\kappa\theta\right)~\dot X^Me_M^A
\nonumber\\&&
+\frac{1}{4}\left(
\bar\theta\Gamma^A\delta_\kappa\theta~
\bar\theta\Gamma_A\sigma\dot\theta
+\bar\theta\Gamma_A\sigma\delta_\kappa\theta~
\bar\theta\Gamma^A\dot\theta
\right)~,\\
\CL^{\mu}&=&\Big[
-\frac{1}{2}\bar\theta\Gamma^C\delta_\kappa\theta~
(\bar\theta\Gamma_A\sigma\varrho_C\theta+\CF_{AB}\bar\theta\Gamma^B\varrho_C\theta)
\nonumber\\&&
+\frac{1}{4}\left(
\bar\theta\Gamma^B\delta_\kappa\theta~
\bar\theta\Gamma_B\sigma\varrho_A\theta
+\bar\theta\Gamma_B\sigma\delta_\kappa\theta~
\bar\theta\Gamma^B\varrho_A\theta
\right)
\nonumber\\&&
-\frac{i}{12}(\bar\theta\Gamma_A\sigma\CM^2\delta_\kappa\theta
+\CF_{AB}\bar\theta\Gamma^B\CM^2\delta_\kappa\theta)
\Big]~\dot X^Me_M^A~,
\\
\CL^{\rm spin}&=&\Big[
-\frac{1}{2}\bar\theta\Gamma^C\delta_\kappa\theta~
(\bar\theta\Gamma_A\sigma\varrho^{\rm spin}_C\theta
+\CF_{AB}\bar\theta\Gamma^B\varrho^{\rm spin}_C\theta)
\nonumber\\&&
+\frac{1}{4}\left(
\bar\theta\Gamma^B\delta_\kappa\theta~
\bar\theta\Gamma_B\sigma\varrho^{\rm spin}_A\theta
+\bar\theta\Gamma_B\sigma\delta_\kappa\theta~
\bar\theta\Gamma^B\varrho^{\rm spin}_A\theta
\right)
\Big]~\dot X^Me_M^A~.
\end{eqnarray}
In the next subsection we will examine
these surface terms and determine boundary conditions 
under which these vanish.

\subsection{1/4 BPS noncommutative D-branes in the pp-wave}

In order to determine the boundary conditions for the fermionic variable
$\theta$\,, we shall introduce the gluing matrix $\BM$ given in
(\ref{M}) and demand that $\theta=\BM\theta$ in the same way as
in the AdS
case.  Since it satisfies (\ref{key})\,, $\CL^0$ vanishes.

By using (\ref{key}), $\CL^{\mu}$ can be rewritten as
\begin{eqnarray}
&&\Big[
\frac{1}{2}
\bar\theta\Gamma^{\bar C}\delta_\kappa\theta~
\bar\theta(\Gamma_{\bar A}
+\CF_{\bar A\bar B}\Gamma^{\bar B})
\BM \varrho_{\bar C}
\theta
\nonumber\\&&
-\frac{1}{4}
\bar\theta\Gamma^{\bar B}\delta_\kappa\theta~
\bar\theta(\Gamma_{\bar B}
+\CF_{\bar B\bar C}
\Gamma^{\bar C})
\BM \varrho_{\bar A}
\theta
\nonumber\\&&
+\frac{1}{4}\bar\theta\Gamma_{\underline{B}}\sigma_3\delta_\kappa\theta~
\bar\theta\Gamma^{\underline{B}}\varrho_{\bar A}\theta
\nonumber\\&&
+\frac{i}{12}
\bar\theta(
\Gamma_{\bar A}+\CF_{\bar A\bar B}\Gamma^{\bar B}
)\BM \CM^2
\delta_\kappa\theta
\Big]~\dot X^Me_M^{\bar A}~
\label{L^mu alternative}
\end{eqnarray}
where we have used
\begin{eqnarray}
\bar\theta\Gamma^{\underline{B}}\delta_\kappa\theta
=\frac{1}{2}\bar\theta(\BM'\Gamma^{\underline{B}}
+\Gamma^{\underline{B}}\BM)
\delta_\kappa\theta
=0~.
\end{eqnarray}
The vanishing condition for the first and the second lines is
\begin{eqnarray}
\BM \varrho_{\bar C}\theta=
\varrho_{\bar C}\BM\theta~,
\label{cond 1 pp}
\end{eqnarray}
which ensures that the third line vanishes.
(\ref{cond 1 pp}) means
\begin{eqnarray}
\BM \varrho_{\bar m}\theta=
\varrho_{\bar m}\BM\theta
\label{mu-1:m}
\end{eqnarray}
for $-\in D$\,,
while for $-\in N$\,,
(\ref{mu-1:m}) and 
\begin{eqnarray}
\BM \varrho_{\bar -}\theta=
\varrho_{\bar -}\BM\theta~.
\label{mu-1:-}
\end{eqnarray}
Among other $p$, $p=1$ is the special case.
For $p=1$ and $-\in D$,
(\ref{mu-1:m}) is satisfied when 
\begin{eqnarray}
\varphi_n\Gamma^{\bar a_{2n-1}\bar a_{2n}}f
=f\varphi_n\Gamma^{\bar a_{2n-1}\bar a_{2n}}~~~\text{and}~~~
d=\text{even}~,
\end{eqnarray}
where $d~(d')$ is the number of Dirichlet directions 
contained in $\{1,2,3,4\}$ ($\{5,6,7,8\}$),
respectively. 
This means that the condition is satisfied by $(2,0)$- and $(0,2)$-branes.
On the other hand, for $-\in N$\,,
(\ref{mu-1:-}) implies that
\begin{eqnarray}
\varphi_n\Gamma^{\bar a_{2n-1}\bar a_{2n}}(f+g)
=-(f+g)\varphi_n\Gamma^{\bar a_{2n-1}\bar a_{2n}}~~~\text{and}~~~
d=\text{even}~,
\end{eqnarray}
which is not satisfied by the $(+-)$-brane.
Thus we find that for (0,2)- and (2,0)-branes the first and the second lines vanish
and so the third line vanishes.
It is straightforward to see that the last line vanish for these branes.
As will be seen below, $\CL^{\rm spin}$ vanishes for these
branes sitting even outside the origin.
Thus we find that (0,2)- and (2,0)-branes sitting anywhere
are 1/2 BPS.

For 1/4 BPS branes\footnote{
We should note  that 1/4 BPS branes we consider here
is a part of general 1/4 BPS branes.
The 1/4 projectors here are composed of two of mutually commuting
gluing matrices only.
}, we
 introduce an additional gluing matrix $\BM_0$ given in (\ref{M_0})
which commutes with $\BM$, $[\BM,\BM_0]=0$,
and satisfies
\begin{eqnarray}
\BM_0'=C^{-1}\BM_0^TC=
(-1)^{p+1+[\frac{p+1}{2}]}\BM_0~.
\end{eqnarray}
First we examine the first and the the second lines
in (\ref{L^mu alternative}).
By using the boundary condition $\theta=\BM_0\theta$,
\begin{eqnarray}
\bar\theta\Gamma^{\bar B}\delta_\kappa\theta=0
~~~~\text{for}~~ p=3\mod 4
\label{N M_0}
\end{eqnarray}
is derived and 
so they vanish for $p=3\mod 4$.
For $p=1\mod 4$,
since
\begin{eqnarray}
\bar\theta\Gamma^{\bar B}\BM \varrho_{\bar C}\theta
=-\bar\theta\BM_0 \Gamma^{\bar B}\BM \varrho_{\bar C}\theta
= \bar\theta\Gamma^{\bar B}\BM \BM_0\varrho_{\bar C}\theta
\end{eqnarray}
and
\begin{eqnarray}
\BM_0\varrho_{\bar -}&=&\frac{\mu}{2}((-1)^df+(-1)^{d'}g)i\sigma_2\BM_0
\nonumber\\
\BM_0\varrho_{\bar m}&=&\left\{
  \begin{array}{ll}
-(-1)^{d}\varrho_{\bar r}\BM_0~,~
-(-1)^{d'}\varrho_{\bar r'}\BM_0~,~~       ~~~~~& -\in D   \\
0       &-\in N     \\
  \end{array}
\right.~,
\label{varrho M_0}
\end{eqnarray}
they vanish when
one of the followings is satisfied
\begin{itemize}
  \item $p=3\mod 4$,
  \item $p=1\mod 4$, $-\in D$ and $d=$ even,
  \item $p=1\mod 4$, $-\in N$ and $d=$ odd.
\end{itemize}

Next, we examine the third line.
It vanishes when $p=1\mod 4$
because
\begin{eqnarray}
\bar\theta\Gamma^{\underline{B}}\sigma_3\delta_\kappa\theta=0
~~~~\text{for}~ p=1\mod 4~.
\label{D M_0}
\end{eqnarray}
For $p=3\mod 4$, since
\begin{eqnarray}
\bar\theta\Gamma^{\underline{B}}\varrho_{\bar A}\theta
=\bar\theta\Gamma^{\underline{B}}\BM_0\varrho_{\bar A}\theta
\end{eqnarray}
and (\ref{varrho M_0}),
the third line vanishes when one of the followings is satisfied
\begin{itemize}
  \item $p=1\mod 4$,
  \item $p=3\mod 4$, $-\in D$ and $d=$ even,
  \item $p=3\mod 4$, $-\in N$ and $d=$ odd.
\end{itemize}

Summarizing we find that the first, second and third lines 
vanish
for ($+-$;odd,odd)-branes and (even,even)-branes.

It is straightforward to see that for these branes
the fourth line vanishes.

Next we examine $\CL^{\rm spin}$,
which is rewritten  by using (\ref{key}) as
\begin{eqnarray}
&&\Big[
\frac{1}{2}
\bar\theta\Gamma^{\bar C}\delta_\kappa\theta~
\bar\theta(\Gamma_{\bar A}\sigma_3
+\CF_{\bar A\bar B}\Gamma^{\bar B}
)\BM \varrho^{\rm spin}_{\bar C}
\theta
\nonumber\\&&
-\frac{1}{4}
\bar\theta\Gamma^{\bar B}\delta_\kappa\theta~
\bar\theta(
\Gamma_{\bar B}\sigma_3
+\CF_{\bar B\bar C}~\Gamma^{\bar C}
)\BM \varrho^{\rm spin}_{\bar A}
\theta
\nonumber\\&&
+\frac{1}{4}\bar\theta\Gamma_{\underline{B}}\sigma_3\delta_\kappa\theta~
\bar\theta\Gamma^{\underline{B}}\varrho^{\rm spin}_{\bar A}\theta
\Big]~\dot X^Me_M^{\bar A}~.
\end{eqnarray}
In the presence of $\CF$ on the D-brane world-volume, 
the spin connection 
$\varrho^{\rm spin}_-dX^-=\omega^{m+}_-dX^-$ 
may develop additional components
$\omega^{\bar A\bar B}_-dX^-$.
Even in this case, 
it is natural to expect that
\begin{eqnarray}
[\varphi_n\Gamma^{\bar a_{2n-1}\bar a_{2n}},
\omega^{\bar A\bar B}\Gamma_{\bar A\bar B}]
=0~
\label{omega}
\end{eqnarray}
as in the AdS case.
For $-\in D$, it vanishes as $\varrho^{\rm spin}_C\neq 0$ only for $C=-$,
so we consider the case with $-\in N$ below.
The first and the second lines vanish for $p=3\mod 4$ due to (\ref{N M_0}).
Even for $p=1\mod 4$, they vanish since
we derive 
\begin{eqnarray}
\bar\theta(\Gamma_{\bar A}\sigma_3+\CF_{\bar A\bar B}\Gamma^{\bar B})
\Gamma_{\bar C\bar D}\theta\omega^{\bar C\bar D}
=-\bar\theta(\Gamma_{\bar A}\sigma_3+\CF_{\bar A\bar B}\Gamma^{\bar B})
\Gamma_{\bar C\bar D}\BM\theta\omega^{\bar C\bar D}
=0~,
\end{eqnarray}
and
\begin{eqnarray}
\bar\theta(\Gamma_{\bar A}\sigma_3+\CF_{\bar A\bar B}\Gamma^{\bar B})
\BM
\frac{\mu^2}{2}X^{\underline{m}}\Gamma_{\underline{m}}\Gamma_+\theta
&=&-\bar\theta\BM_0(\Gamma_{\bar A}\sigma_3+\CF_{\bar A\bar B}\Gamma^{\bar B})
\BM
\frac{\mu^2}{2}X^{\underline{m}}\Gamma_{\underline{m}}\Gamma_+\theta
\nonumber\\
&=&0~.
\end{eqnarray}
Next we examine the last line,
which vanishes
when  $p=1\mod 4$ due to (\ref{D M_0}).
For $p=3\mod 4$,
it vanishes since
we derive assuming (\ref{omega})
\begin{eqnarray}
\bar\theta\Gamma^{\underline{B}}
\Gamma_{\bar C\bar D}\theta\omega^{\bar C\bar D}
=-\bar\theta\Gamma^{\underline{B}}
\Gamma_{\bar C\bar D}\BM\theta\omega^{\bar C\bar D}
=0~,
\end{eqnarray}
and
\begin{eqnarray}
\bar\theta\Gamma^{\underline{B}}
\frac{\mu^2}{2}X^{\underline{m}}\Gamma_{\underline{m}}\Gamma_+\theta
=\bar\theta\Gamma^{\underline{B}}\BM_0
\frac{\mu^2}{2}X^{\underline{m}}\Gamma_{\underline{m}}\Gamma_+\theta
=0~.
\end{eqnarray}
Summarizing $\CL^{\rm spin}$ vanishes for ($+-$;odd,odd)-branes and
(even,even)-branes even if they are sitting outside the origin.
Finally, it is obvious that (0,2)- and (2,0)-branes are 1/2 BPS even off
the origin, because $-\in D$ for these branes.  As a result, we have
obtained noncommutative D-branes summarized in Table \ref{1/4 D pp-wave}.

\vspace*{0.5cm}
\begin{table}[htbp]
 \begin{center}
  \begin{tabular}{|c|c|c|c|c|c|}
    \hline
SUSY& D1      & D3   &D5    & D7   & D9   \\
    \hline
1/2&(0,2),~(2,0)       &   & &   &absent \\
1/4&      &(4,0),~(2,2),~(0,4)    &(4,2),~(2,4)    &(4,4)    &\\
&       & ($+-$;1,1)   &($+-$;3,1),~($+-$;1,3)     &($+-$;3,3)     &    \\
    \hline
  \end{tabular}
\caption{\Nc D-branes in the pp-wave}
\label{1/4 D pp-wave}
 \end{center}
\end{table}

In the commutative limit $\forall\CF\to 0$,
\nc D-branes above are reduced to commutative D-branes.
Combining the result obtained in the flat limit
with that obtained in \cite{Sakaguchi:2003py},
we summarize commutative D-branes sitting at the origin 
of the pp-wave in Table \ref{C D pp-wave}. 

\vspace*{0.5cm}
\begin{table}[htbp]
 \begin{center}
  \begin{tabular}{|c|c|c|c|c|c|c|}
    \hline
SUSY&D($-1$)& D1      & D3   &D5    & D7   & D9   \\
    \hline
1/2&(0,0)&(0,2),~(2,0)       &(1,3),~(3,1)   &(2,4),~(4,2) &   &absent \\
&&      &($+-$;0,2)  &($+-$;1,3) &($+-$;2,4) & \\
&&      &($+-$;2,0)   &($+-$;3,1)  &($+-$;4,2)    & \\
1/4&&      &(4,0),~(2,2),~(0,4)    &   &(4,4)    &\\
&&       & ($+-$;1,1)   &    &($+-$;3,3)     &    \\
    \hline
  \end{tabular}
\caption{Commutative D-branes in the pp-wave (sitting at the origin)}
\label{C D pp-wave}
 \end{center}
\end{table}

\section{Summary and Discussions}

We have considered some possible configurations of 1/4 BPS
noncommutative D-branes in the AdS$_5\times$S$^5$ and the pp-wave. The
1/4 BPS noncommutative AdS-branes 
are allowed to exist at arbitrary position in the spacetime. The
D-string case is exceptional and it is 1/2 BPS at the origin and 1/4 BPS
outside the origin. We also have seen that all of them are reduced to
1/2 BPS and 1/4 BPS commutative AdS-branes in the commutative limit or
the strong magnetic flux limit. The 1/4 BPS noncommutative D-branes in
the pp-wave background have also been classified by applying the same
analysis as in the AdS case to the pp-wave case. The resulting possible
D-branes are consistently reproduced from the noncommutative AdS-branes
via the Penrose limit.

It would be possible to see that the result shown in this paper is still
valid at full order in $\theta$ 
by following our previous paper
\cite{Sakaguchi:2004md}, though the possible noncommutative D-branes
have been classified at fourth order in $\theta$\,. This issue is more
complicated and so we leave it as a future problem. We hope that we
could report on this issue in the near future.

It is also interesting to consider the interpretation of our results
before taking a near-horizon limit. According to the work of Skenderis
and Taylor \cite{SMT}, an AdS-brane is surely related to a
supersymmetric intersection of D-brane with a stack of $N$
D3-branes. Hence our result may also be interpreted in terms of the
intersecting D-branes. In particular, the T-dual picture would be
related to an intersecting D-branes at angles
\cite{intersection}. For this direction, it would also be useful to
generalize the work \cite{SMT,AR} by including a constant
two-form. There the relation to dCFTs is also discussed. As a
generalization of our work, it would be interesting to consider an
intersecting AdS D-branes (for an intersecting D-branes on a pp-wave,
see \cite{Ohta}), though we have discussed a single AdS D-brane
here.

As another generalization of our works, it would be interesting to 
consider oblique D-branes \cite{oblique} with gauge field condensates
in the pp-wave background as discussed in \cite{Mattik,BH-Lee}. 
It would also be nice to study the AdS origin of the oblique D-branes 
by following \cite{Zamaklar}. 
 
As an interesting application of our procedure, we can consider
noncommutative M-branes. For noncommutative M-branes, the classification
of the possible Dirichlet branes is drastically modified and one can see
some interesting features. For this subject we will report in other
papers soon \cite{SY:NCM(flat),SY:NCM(AdS)}.

\section*{Acknowledgments}

The authors thank Y.~Hikida and Y.~Susaki for useful discussions.  This
work is supported in part by the Grant-in-Aid for Scientific Research
(No.~17540262 and No.~17540091) from the Ministry of Education, Science
and Culture, Japan. The work of K.~Y.\ is supported in part by JSPS
Research Fellowships for Young Scientists.

\end{document}